\documentstyle[12pt,epsfig]{article}
\def\lbldef#1#2{\expandafter\gdef\csname #1\endcsname {#2}}

\def\href#1#2{#2}  

\begin{document}
\baselineskip=15.5pt
\pagestyle{plain}
\setcounter{page}{1}

\begin{titlepage}

\begin{flushright}
CERN-TH/2000-157\\
hep-th/0006019
\end{flushright}
\vspace{10 mm}

\begin{center}
{\Large A Note on (D$(p-2)$,D$p$) Bound State and Noncommutative 
Yang-Mills Theory}

\vspace{5mm}

\end{center}

\vspace{5 mm}

\begin{center}
{\large Donam Youm\footnote{E-mail: Donam.Youm@cern.ch}}

\vspace{3mm}

Theory Division, CERN, CH-1211, Geneva 23, Switzerland

\end{center}

\vspace{1cm}

\begin{center}
{\large Abstract}
\end{center}

\noindent

We give a microscopic explanation for the recently observed equivalence 
among thermodynamics of supergravity solutions for D$p$-branes with 
or without NS $B$-field and for D$(p-2)$-branes with vanishing 
$B$-field and two delocalized transverse directions by showing that 
these D-brane configurations are related to one another through $T$-duality 
transformations.  This result also gives an evidence for the equivalence among 
the noncommutative and the ordinary Yang-Mills theories corresponding to the 
decoupling limits of the worldvolume theories of such D-brane configurations.

\vspace{1cm}
\begin{flushleft}
CERN-TH/2000-157\\
June, 2000
\end{flushleft}
\end{titlepage}
\newpage

\section{Introduction}

Recently, much attention has been given to noncommutative Yang-Mills 
theory (NCYM), after it is realized that NCYM arises naturally in a 
specific compactification of the matrix theory \cite{cds} and in string 
theories as decoupling limits of the worldvolume theories on D-branes in 
nonzero $B$-field background \cite{dh,sw}.  It is argued \cite{sw} that 
NCYM and the ordinary Yang-Mills theory (CYM) arise from the same field 
theory regularized in different ways.  The gauge fields of NCYM and CYM 
are related \cite{sw} by requiring the equivalence of the gauge 
transformations of the two theories.  The evidence for the equivalence 
between the large $N$ limit NCYM and CYM was also given in Refs. 
\cite{bs,av}.  

In the spirit of the holographic principle \cite{hoo,sus1,sus2}, the bulk 
supergravity dual description of NCYM was studied in Refs. \cite{hi,mar,aos}.  
In accordance with the conjectured equivalence between NCYM and CYM, 
it is found out that thermodynamics of near-extremal D$p$-branes with 
nonzero $B$-field coincides exactly with that of the 
corresponding D$p$-brane without $B$-field \cite{mar,aos,br,co1,ho}.  
By observing that the near horizon geometry and thermodynamics of the 
supergravity solution for the D$p$-brane with nonzero $B$-field 
are identical to those of the supergravity solution for the 
D$(p-2)$-brane with vanishing $B$-field and two of its transverse 
coordinates delocalized, it is further argued \cite{lur,co2} that NCYM 
in $p+1$ dimensions and CYM in $p-1$ dimensions are equivalent.

It is the purpose of this paper to give a microscopic explanation for 
the equivalence of thermodynamics of various supergravity solutions 
for the D-brane systems mentioned in the above.  We relate such supergravity 
solutions to microscopic D-brane systems through the boundary state formalism 
and show that such D-brane systems are related to one another through 
$T$-duality transformations, implying that the number of microscopic degrees 
of freedoms of the above mentioned different D-brane configurations are 
mapped one-to-one to one another under the $T$-duality transformations.  
We begin by reviewing the relevant ideas and elaborating on their connection 
in section 2, for the purpose of preparing for the discussion of the main 
result of this paper.  In section 3, we show that various D-brane systems 
whose thermodynamics are shown to coincide with one another are related under 
the $T$-duality transformations.  

\section{General Properties}

The bosonic action for an open string ending on a D$p$-brane in the 
constant NS $B$-field background can be written in the following form:
\begin{equation}
S={1\over{4\pi\alpha^{\prime}}}\int d^2\xi\left[\sqrt{-h}
h^{\alpha\beta}g_{\mu\nu}\partial_{\alpha}X^{\mu}\partial_{\beta}X^{\nu}
-\epsilon^{\alpha\beta}{\cal F}_{ij}\partial_{\alpha}X^i\partial_{\beta}
X^j\right],
\label{ndpact}
\end{equation}
by going to the gauge in which the NS $B$-field takes the form $B=B_{ij}
dX^i\wedge dX^j$ ($i,j=0,...,p$) \cite{wit}.  Here, ${\cal F}_{ij}\equiv 
B_{ij}+\partial_{[i}A_{j]}$, $A_i$ is the $U(1)$ gauge field living on the 
D$p$-brane worldvolume and $\xi^{\alpha}=(\tau,\sigma)$ is the string 
worldsheet coordinates.  By applying the gauge transformation 
$B_{ij}\to B_{ij}+\partial_i\Lambda_j-\partial_j\Lambda_i$ and 
$A_i\to A_i-\Lambda_i$, under which the action (\ref{ndpact}) is invariant, 
one can set $A_i$ equal to zero, which we assume from now on. 
In the $h_{\alpha\beta}=\eta_{\alpha\beta}$ gauge with constant 
$g_{\mu\nu}$ and $B_{ij}$, the variation of the bosonic open string action 
(\ref{ndpact}) with respect to the string coordinates $X^{\mu}$ yields the 
the following boundary conditions at the ends $\sigma=0,\pi$ of the open 
string:
\begin{eqnarray}
g_{ij}\partial_{\sigma}X^j+B_{ij}\partial_{\tau}X^j=0,\ \ \ \ \ 
i,j=0,1,...,p,
\cr
\delta X^a=0,\ \ \ \ \ a=p+1,...,9.
\label{bdrycdtn}
\end{eqnarray}
The string propagator with this boundary conditions takes the following form 
\cite{ft,cln,sw}:
\begin{equation}
\langle X^i(z)X^j(z^{\prime})\rangle=-\alpha^{\prime}\left[
g^{ij}\ln{{|z-z^{\prime}|}\over{|z-\bar{z}^{\prime}|}}+G^{ij}\ln
|z-\bar{z}^{\prime}|^2+{1\over{2\pi\alpha^{\prime}}}\theta^{ij}
\ln{{z-\bar{z}^{\prime}}\over{\bar{z}-z^{\prime}}}\right],
\label{propagator}
\end{equation}
where $z=\tau+i\sigma$ and
\begin{eqnarray}
G^{ij}&=&\left({1\over{g+B}}\right)^{ij}_S=
\left({1\over{g+B}}g{1\over{g-B}}\right)^{ij},
\cr
\theta^{ij}&=&2\pi\alpha^{\prime}\left({1\over{g+B}}\right)^{ij}_A=
-2\pi\alpha^{\prime}\left({1\over{g+B}}B{1\over{g-B}}\right)^{ij},
\label{propdefs}
\end{eqnarray}
where the subscripts $S$ and $A$ respectively denote the symmetric and 
the antisymmetric parts of the matrix.  
$G^{ij}$ is interpreted as the effective metric seen by the open strings and 
$\theta^{ij}$ has the interpretation as the noncommutativity parameter as can 
be seen from the following time ordered commutation relation that follows from 
Eq. (\ref{propagator}):
\begin{equation}
[X^i(\tau),X^j(\tau)]=T\left(X^i(\tau)X^j(\tau^-)-X^i(\tau)X^j(\tau^+)\right)
=i\theta^{ij}.
\label{noncommrel}
\end{equation}
So, the end-points of the open string target space coordinates associated 
with non-zero components of the constant $B$ field live in noncommutative 
space.  Namely, in the presence of nonzero $B$-field, the D-brane 
worldvolume becomes noncommutative.  This result was also obtained by the 
quantization of $X^i$ through the analysis of the time-averaged symplectic 
form on the phase space \cite{ch} or through the Dirac bracket quantization 
procedure \cite{aas}.  (See also Refs. \cite{sj1,sj2,sj3}.)  

D-branes can be alternatively described by the ``boundary states'' 
\cite{cln,pc} of the ``close-string channel'' description.  The 
``open-string channel'' and the ``closed-string channel'' descriptions of 
D-branes are mapped to one another through $\tau\leftrightarrow\sigma$, 
under which an open string one-loop diagram and a closed string tree diagram 
are interchanged.   The boundary state is given by the product of a matter 
and a ghost parts, each of which is expressed as the product of the bosonic 
and the fermionic parts.  The GSO projection selects specific linear 
combinations of such boundary states separately for the NS-NS and the R-R 
sectors.  In this paper, we will be mainly concerned with the bosonic matter 
part $|B_X\rangle$ of the boundary state.  By applying the transformation 
$\tau\leftrightarrow\sigma$ (followed by an appropriate rescaling of 
$\tau$ and $\sigma$) to the boundary conditions (\ref{bdrycdtn}) on an open 
string, one obtains the following conditions on $|B_X\rangle$ at $\tau=0$:
\begin{eqnarray}
(\partial_{\tau}X^i+B^i_{\ j}\partial_{\sigma}X^j)|_{\tau=0}
|B_X\rangle=0,\ \ \ \ \ i,j=0,1,...,p,
\cr
(X^a|_{\tau=0}-x^a)|B_X\rangle=0,\ \ \ \ \ a=p+1,...,9.
\label{bosmattbdcond}
\end{eqnarray}
In the nonzero $B$-field background with some of coordinates compactified on 
a torus, the oscillator expansion for the closed string target space 
coordinates is
\begin{equation}
X^{\mu}=x^{\mu}+w^{\mu}\sigma+\tau g^{\mu\nu}(p_{\nu}-B_{\nu\rho}w^{\rho})
+{i\over\sqrt{2}}\sum_{n\neq 0}\left[{{\alpha^{\mu}_n}\over{n}}
e^{-in(\tau-\sigma)}+{{\tilde{\alpha}^{\mu}_n}\over{n}}e^{-in(\tau+\sigma)}
\right],
\label{cstrexp}
\end{equation}
where $w^{\mu}$ is zero for noncompact directions and we assume that only 
the longitudinal components $B_{ij}$ of the two-form potential are nonzero.  
So, the conditions (\ref{bosmattbdcond}) on $|B_X\rangle$ at $\tau=0$ in terms 
of the oscillator modes take the following forms:
\begin{eqnarray}
& &\hat{p}^i|B_X\rangle=0,\ \ \ \ \ \ \ \ 
(\hat{x}^a-x^a)|B_X\rangle=0,
\cr
& &\left[({\bf 1}+B)^i_{\ j}\alpha^j_n+({\bf 1}-B)^i_{\ j}
\tilde{\alpha}^j_n\right]|B_X\rangle=0,
\cr
& &(\alpha^a_n-\tilde{\alpha}^a_{-n})|B_X\rangle=0,\ \ \ \ \ \ 
\hat{w}^a|B_X\rangle=0,
\label{bsmatoscon}
\end{eqnarray}
where $\bf 1$ is the $(p+1)\times(p+1)$ identity matrix.

We now discuss the $T$-duality transformation \cite{giv} of closed string 
theory on $T^d$ and its effect on D-brane configurations.  The $T^d$ part 
of the canonical Hamiltonian of the closed string is
\begin{equation}
H={1\over{4\pi\alpha^{\prime}}}\int^{2\pi}_0d\sigma\,
\left(\matrix{X^{\prime}& 2\pi\alpha^{\prime}P}\right) M(E)
\left(\matrix{X^{\prime}\cr 2\pi\alpha^{\prime}P}\right),
\label{clstrham}
\end{equation}
where the matrix $M(E)$ determined by $E=g+B$ is given by
\begin{equation}
M(E)=\left(\matrix{g-Bg^{-1}B& Bg^{-1}\cr -g^{-1}B& g^{-1}}\right),
\label{matrixe}
\end{equation}
$X^{\prime}=\partial_{\sigma}X$ and $P=(g\partial_{\tau}X+B
\partial_{\sigma}X)/(2\pi\alpha^{\prime})$ is the conjugate momentum.  
It appears from Eq. (\ref{clstrham}) that the Hamiltonian has the 
$O(d,d,{\bf R})$ symmetry, but since the eigenvalues of the operators 
$\hat{w}^i$ and $\hat{p}^i$ in the mode expansion (\ref{cstrexp}) take 
integer values due to the periodicity condition $X^i\sim X^i+2\pi$ of the 
compactified coordinates actually the Hamiltonian is invariant only under 
the $O(d,d,{\bf Z})$ subset
\footnote{The Hamiltonian for the open string has the same form 
(\ref{clstrham}) in terms of the string coordinates and the conjugate 
momentum and therefore appears to have the same symmetry as the closed 
string case.  However, the $O(d,d,{\bf Z})$ target space duality symmetry 
is not a symmetry of the open string due to the absence of the winding 
modes.}.  

Under the $O(d,d,{\bf Z})$ transformation with the transformation matrix 
$\left(\matrix{a&b\cr c&d}\right)\in O(d,d,{\bf Z})$, the background fields 
and the oscillator modes transform as
\begin{eqnarray}
E&\to& E^{\prime}=(aE+b)(cE+d)^{-1},
\cr  
\alpha_n(E)&\to&(d-cE^T)^{-1}\alpha_n(E^{\prime}),\ \ \  
\tilde{\alpha}_n(E)\to (d+cE)^{-1}\tilde{\alpha}_n(E^{\prime}).
\label{oddtran}
\end{eqnarray}
Note, this transformation is valid also for the $n=0$ case, where 
the oscillator modes are defined as
\begin{equation}
\alpha_0(E)\equiv {1\over\sqrt{2}}G^{-1}(p-Ew),\ \ \ \ 
\tilde{\alpha}_0(E)\equiv {1\over\sqrt{2}}G^{-1}(p+E^Tw).
\label{zeroscilmod}
\end{equation}

By making use of the equivalence between the closed-string and open-string 
channel descriptions of D-branes through boundary states, we study the effect 
of $T$-duality symmetry of closed string theory on D-branes.  For this 
purpose, it is convenient to divide the closed string coordinate mode 
expansion of the form (\ref{cstrexp}) into the left-moving and the 
right-moving parts as $X={1\over 2}(X_-+X_+)$ with
\begin{eqnarray}
X_-&=&x+\sqrt{2}(\tau-\sigma)\alpha_0+i\sqrt{2}\sum_{n\neq 0}{{\alpha_n}
\over n}e^{-in(\tau-\sigma)},
\cr
X_+&=&x+\sqrt{2}(\tau+\sigma)\tilde{\alpha}_0+i\sqrt{2}\sum_{n\neq 0}
{{\tilde{\alpha}_n}\over n}e^{-in(\tau+\sigma)},
\label{pmclstrcord}
\end{eqnarray}
where the zero modes $\alpha_0$ and $\tilde{\alpha}_0$ are defined in Eq. 
(\ref{zeroscilmod}).  The effect of the $O(d,d,{\bf Z})$ symmetry 
transformation of the closed string theory on the open-string channel can 
be inferred by noting the map $\tau\leftrightarrow\sigma$ (and therefore 
$\partial_{\tau}\leftrightarrow\partial_{\sigma}$) that connects closed 
and open string channel descriptions of D-branes.  The $O(d,d,{\bf Z})$ 
$T$-duality transformation (\ref{oddtran}) is generated by the following 
transformations:
\begin{itemize}
\item Factorized dualities $D_i$:
\begin{equation}
\left(\matrix{a&b\cr c&d}\right)=\left(\matrix{I_d-e_i&e_i\cr e_i&I_d-e_i}
\right),
\label{factdual}
\end{equation}
where $I_d$ is the $d\times d$ identity matrix and the $d\times d$ matrix 
$e_i$ has zero entries except for the $(i,i)$-component which is 1.  
When the $B$-field is zero and $g$ is diagonal, i.e. $g={\rm diag}
(R^2_1,...,R^2_d)$, the factorized duality $D_i$ acts as the well-known 
large and small radii duality on the $i$-th coordinate, i.e. $R_i\to 
1/R_i$ while the remaining radii unchanged.  The oscillator modes transform
\footnote{The extra factor of $1/R^2_i$ in the oscillator mode 
transformation is due to our choice of gauge in which the periodicity 
of the coordinates $X$ is fixed to be $2\pi$ and all the information 
on the size and the shape of the torus is encoded in the background 
field $E=g+B$.  We have chosen such gauge because we wish to consider 
the general constant background fields $g$ and $B$.}
as $\alpha^i_n\to -{1\over{R^2_i}}\alpha^i_n$ and $\tilde{\alpha}^i_n\to 
{1\over{R^2_i}}\tilde{\alpha}^i_n$, meanwhile the remaining modes 
$\alpha^j_n$ and $\tilde{\alpha}^j_n$ ($j\neq i$) remain unchanged.  
This implies the transformation $\partial_{\tau}X^i\leftrightarrow
\partial_{\sigma}X^i$ on the $i$-th closed string coordinate $X^i$.   
From the correspondence between the open string boundary conditions 
(\ref{bdrycdtn}) and the conditions (\ref{bosmattbdcond}) on the boundary 
states, one can see therefore that the Dirichlet and the Neumann boundary 
conditions of the $i$-th open string coordinate are interchanged.  For a 
general background $E=g+B$, although $E$ transforms in more complicated way, 
the oscillator modes transform similarly as the diagonal $E$ case, namely 
$\alpha^i_n\to -{1\over{g_{ii}}}\alpha^i_n$ and $\tilde{\alpha}^i_n\to 
{1\over{g_{ii}}}\tilde{\alpha}^i_n$ with the remaining modes unchanged, and 
therefore $\partial_{\tau}X^i\leftrightarrow\partial_{\sigma}X^i$.  
So, the factorized dualities generally correspond to the usual $T$-duality 
transformations that transform D$p$-brane into D$(p\pm 1)$-brane.

\item Basis change of the compactification lattice $\Lambda$, i.e. 
$E\to AEA^T$ with $A\in GL(d,{\bf Z})$:
\begin{equation}
\left(\matrix{a&b\cr c&d}\right)=\left(\matrix{A&0\cr 0&(A^T)^{-1}}\right)
\ \ \ {\rm s.t.}\ \ \ A\in GL(d,{\bf Z}).
\label{bssch}
\end{equation}
Under this transformation, the background fields and the oscillator modes 
transform as
\begin{equation}
g\to AgA^T,\ \ \ B\to ABA^T,\ \ \ 
\alpha_n(E)\to A^T\alpha_n(E^{\prime}),\ \ \ 
\tilde{\alpha}_n(E)\to A^T\tilde{\alpha}_n(E^{\prime}).
\label{bschtrns}
\end{equation}
So, the derivatives of closed string coordinates transform as
\begin{equation}
\partial_{\tau}X\to A^T\partial_{\tau}X,\ \ \ \ \ 
\partial_{\sigma}X\to A^T\partial_{\sigma}X,
\label{bschcrdtrn}
\end{equation}
implying that the worldvolume dimensionality of a D-brane does not change 
under the $GL(d,{\bf Z})$ transformation.  We see therefore that the 
D-brane system with the constant $g$ and $B$ and the associated NCYM are 
equivalent to those with $AgA^T$ and $ABA^T$, where $A\in GL(d,{\bf Z})$.  

\item Integer ``$\Theta$''-parameter shift of $E$, i.e. $E_{ij}\to E_{ij}
+\Theta_{ij}$ with $\Theta_{ij}=-\Theta_{ji}\in {\bf Z}$:
\begin{equation}
\left(\matrix{a&b\cr c&d}\right)=\left(\matrix{I_d&\Theta\cr 0&I_d}\right)
\ \ \ {\rm s.t.}\ \ \ \Theta^T=-\Theta.
\label{thetatran}
\end{equation}
Under this symmetry, the background fields and the oscillator modes 
transform as
\begin{equation}
g\to g,\ \ \ B\to B+\Theta,\ \ \ 
\alpha_n(E)\to\alpha_n(E^{\prime}),\ \ \ 
\tilde{\alpha}_n(E)\to\tilde{\alpha}_n(E^{\prime}).
\label{thetabkosc}
\end{equation}
implying that closed string coordinates remain unchanged in the form 
$X={1\over 2}(X_-+X_+)$, meanwhile the $B$-field shifts by the integer-valued 
antisymmetric matrix $\Theta$ as $B\to B+\Theta$.  So, NCYM associated with 
D$p$-brane with constant $B$ is equivalent to NCYM associated with D$p$-brane 
with $B+\Theta$.  In particular, CYM associated with D$p$-brane without 
$B$-field is equivalent to NCYM associated with D$p$-brane in the 
integer-valued $B$-field background.

\end{itemize}

\section{D$p$-Brane with Rank 2 $B$ Field and (D$(p-2)$,D$p$) Bound State}

In this section, we restrict our attention to the case of D$p$-brane with 
the rank 2 NS $B$-field and $g_{\mu\nu}=\eta_{\mu\nu}$.  We choose the 
non-zero component of the $B$-field to be $B_{p-1,p}$.  In this case, the 
boundary condition (\ref{bdrycdtn}) for the open string takes the following 
form:
\begin{eqnarray}
\partial_{\sigma}X^i&=&0,\ \ \ \ \ \ i=0,1,...,p-2,
\cr
\partial_{\sigma}X^{i^{\prime}}+B_{i^{\prime}j^{\prime}}\partial_{\tau}
X^{j^{\prime}}&=&0,\ \ \ \ \ \ i^{\prime},j^{\prime}=p-1,p,
\cr
\delta X^a&=&0,\ \ \ \ \ \ a=p+1,...,9.
\label{bdrcndpp2}
\end{eqnarray}
In the corresponding closed-string channel description of such 
D-brane configuration, the condition on the bosonic matter boundary state 
at $\tau=0$ is 
\begin{eqnarray}
\partial_{\tau}X^i|_{\tau=0}|B_X\rangle&=&0,\ \ \ \ \ \ i=0,1,...,p-2,
\cr
(\partial_{\tau}X^{i^{\prime}}+B_{i^{\prime}j^{\prime}}\partial_{\sigma}
X^{j^{\prime}})|_{\tau=0}|B_X\rangle&=&0,\ \ \ \ \ \ i^{\prime},j^{\prime}
=p-1,p,
\cr
(X^a|_{\tau=0}-x^a)|B_X\rangle&=&0,\ \ \ \ \ \ a=p+1,...,9.
\label{cntinbdr}
\end{eqnarray}
The boundary condition of the form (\ref{bdrcndpp2}) can also be achieved 
by rotating a D$(p-1)$-brane in the plane defined by one of its longitudinal 
direction (say, $X^{p-1}$) and one of its transverse direction (say, $X^p$) 
and then applying the $T$-duality transformation along the $X^p$-direction.   
The rotation angle $\theta$ is then related to the $B$-field as $B_{p-1,p}=
\tan\theta$.  

To obtain the long distance behavior of the massless bosonic fields in  
closed string states interacting with the D-brane, we project the boundary 
state onto the massless bosonic states \cite{bsg1,bsg2}.  Namely, the long 
distance fluctuation of a field $\Psi$ in the closed string spectrum is 
given by $\delta\Psi=\langle P^{(\Psi)}|D|B\rangle_{\rm NS,RR}$, where 
$P^{(\Psi)}$ is the projector for the field $\Psi$, $D={{\alpha^{\prime}}
\over{4\pi}}\int_{|z|\leq 1}{{d^2z}\over{|z|^2}}z^{L_0-a}
\tilde{z}^{\tilde{L}_0-a}$ ($a=1/2$ in the NS-NS sector and $a=0$ in the 
R-R sector) is the closed string propagator and $|B\rangle_{\rm NS,RR}$ 
is the boundary state for the NS-NS or the R-R sector of the closed string.  
The resulting long distance fluctuation behavior of the massless bosonic 
fields is 
\begin{eqnarray}
\delta\phi&=&{{3-p+2\sin^2\theta}\over{2\sqrt{2}}}nT_p{{V_{p+1}}\over
{k^2_{\perp}}},
\cr
\delta h_{\mu\nu}&=&nT_p{{V_{p+1}}\over{k^2_{\perp}}}{\rm diag}
(-{\sf A},{\sf A},...,{\sf A},{\sf B},{\sf B},{\sf C},...,{\sf C}),
\cr
\delta B_{p-1,p}&=&{{\sin 2\theta}\over\sqrt{2}}nT_p{{V_{p+1}}\over
{k^2_{\perp}}},
\cr
\delta A^{(p-1)}_{01...p-2}&=&\pm\sqrt{2}nT_p\sin\theta{{V_{p+1}}\over
{k^2_{\perp}}},
\cr
\delta A^{(p+1)}_{01...p}&=&\pm\sqrt{2}nT_p\cos\theta{{V_{p+1}}\over
{k^2_{\perp}}},
\label{longdstflds}
\end{eqnarray}
where $T_p=\sqrt{\pi}(2\pi\sqrt{\alpha^{\prime}})^{6-p}$ is the 
tension of a D$p$-brane with the unit brane charge, $V_{p+1}$ is the volume 
of the D$p$-brane worldvolume, $k^2_{\perp}=\sum^9_{a=p+1}k^2_a$ is the 
square of the transverse momentum and 
\begin{equation}
{\sf A}=(p-7-2\sin^2\theta)/8,\ \ \ 
{\sf B}=(p-7+6\sin^2\theta)/8,\ \ \ 
{\sf C}=(p+1-2\sin^2\theta)/8.
\label{defmetcomp}
\end{equation}
Here, we multiplied the entire boundary state by an overall factor 
$n$, which has to be an integer due to the Dirac quantization 
condition, so that the D-brane can take arbitrary R-R charge.  
To express the long distance behavior of the massless fields in ordinary 
space, rather than in momentum space, we apply the following Fourier 
transformation, valid for $p<D-3$:
\begin{equation}
\int d^{p+1}xd^{D-p-1}y{{e^{ik_{\perp}\cdot y}}\over{(D-p-3)r^{D-p-3}
\Omega_{D-p-2}}}={{V_{p+1}}\over{k^2_{\perp}}},
\label{fourier}
\end{equation}
where $r=\sqrt{y^ay^a}$ is the radial coordinate of the (overall) transverse 
space and $\Omega_n=2\pi^{(n+1)/2}/\Gamma((n+1)/2)$ denotes the 
area of a unit $n$-sphere $S^n$.  And then one has to rescale the fields 
in the following way so that the fields can be canonically normalized:
\begin{equation}
\varphi=\sqrt{2}\kappa\phi,\ \ \ 
g_{\mu\nu}=2\kappa h_{\mu\nu},\ \ \ 
{\cal B}_{\mu\nu}=\sqrt{2}\kappa e^{\varphi/2}B_{\mu\nu},\ \ \ 
{\cal A}=\sqrt{2}\kappa A,
\label{rescalfields}
\end{equation}
where $\kappa$ is the ten-dimensional gravitational constant and 
$A$ denotes $A^{(p-1)}$ or $A^{(p+1)}$.  
The resulting long distance behavior of the massless fields exactly 
reproduces the asymptotic behavior of the following Einstein-frame 
supergravity solution for the non-threshold D$p$- and D$(p-2)$-brane bound 
state constructed in Ref. \cite{bmm}:
\begin{eqnarray}
g_{\mu\nu}dx^{\mu}dx^{\nu}&=&H^{-{{7-p}\over 8}}h^{-{1\over 4}}
\left[-dt^2+\cdots+dx^2_{p-2}+h(dx^2_{p-1}+dx^2_p)\right]
\cr
& &+H^{{p+1}\over 8}h^{-{1\over 4}}\left[dy^2_1+\cdots+dy^2_{7-p}\right],
\cr
e^{2\varphi}&=&H^{{3-p}\over 2}h ,\ \ \ \ \ \ \ 
{\cal B}=(1-H^{-1})h\cos\theta\sin\theta\,dx^{p-1}\wedge dx^p,
\cr
{\cal A}^{(p-1)}&=&\pm(1-H^{-1})\sin\theta\,dt\wedge\cdots\wedge dx^{p-2},
\cr 
{\cal A}^{(p+1)}&=&\pm(1-H^{-1})h\cos\theta\,dt\wedge\cdots\wedge dx^{p},
\label{sgsol}
\end{eqnarray}
where $H=1+{{2\kappa nT_p}\over{(7-p)\Omega_{8-p}}}{1\over{r^{7-p}}}$  
and $h^{-1}=\cos^2\theta+H^{-1}\sin^2\theta$.  So, we see that the 
D$p$-brane with the rank 2 constant $B$-field is described in long distance 
region by the D$p$- and D$(p-2)$-brane bound state.  

Just by considering the boundary condition (\ref{bdrcndpp2}) on the open 
string coordinates, it appears that the $B$-field component $B_{p-1,p}$ 
or the rotation angle $\theta$ can take an arbitrary value.  However, it 
turns out that $B_{p-1,p}=\tan\theta$ can take only discrete values 
determined by the Dirac quantization condition for the brane charge 
(density).  From Eq. (\ref{longdstflds}), one can see that the charge 
densities for the D$(p-2)$- and the D$p$-branes are $q_{p-2}=\pm\sqrt{2}
V_2nT_p\sin\theta$ and $q_p=\pm\sqrt{2}nT_p\cos\theta$, where $V_2$ is the 
volume of the $(x_{p-1},x_p)$-plane, which is transverse to the 
D$(p-2)$-branes but is longitudinal to the D$p$-branes.  Since the charge 
density of D$p$-brane takes only values which are integer multiples of the 
fundamental D$p$-brane charge density given by $\mu_p=\sqrt{2}\sqrt{\pi}
(2\pi\sqrt{\alpha^{\prime}})^{3-p}$, we see that $q_p=N_p\sqrt{2}\sqrt{\pi}
(2\pi\sqrt{\alpha^{\prime}})^{3-p}$, where $N_p\in{\bf Z}$ is the total number 
of D$p$-branes.  We note that the relative transverse directions 
$x_{p-1}$ and $x_p$, which are transverse to the D$(p-2)$-branes, are 
delocalized and therefore there are infinitely many fundamental 
D$(p-2)$-branes (with the charge density $\mu_{p-2}=\sqrt{2}\sqrt{\pi}
(2\pi\sqrt{\alpha^{\prime}})^{1-p}$) packed on the $(x_{p-1},x_p)$-plane.  
However, there are finite numbers (say $n_{p-2}$) of D$(p-2)$-branes per 
$(2\pi\sqrt{\alpha^{\prime}})^2$ area of this 2-plane (Cf. Refs. 
\cite{lr1,lr2}).  So, the total number of D$(p-2)$-branes is $N_{p-2}=
n_{p-2}V_2/(2\pi\sqrt{\alpha^{\prime}})^2$ and the total charge 
density is $q_{p-2}=N_{p-2}\mu_{p-2}$.  (The total number $N_{p-2}$ of 
D$(p-2)$-branes is finite [infinite], if the volume $V_2$ of the 2-plane is 
finite [infinite].)  Making use of these facts, we see that the allowed 
values of the $B$-field are restricted by the Dirac quantization condition as
\begin{equation}
B_{p-1,p}=\tan\theta={{q_{p-2}}\over{q_{p}}}{1\over{V_2}}=
{{n_{p-2}}\over{N_{p}}}\ \ \ \ (n_{p-2},N_p\in{\bf Z}).
\label{quantibfield}
\end{equation}
Note, this is valid whether the volume $V_2$ is finite or infinite.  

Since the (rank 2) $B$-field can take only rational values, one can always 
set its non-zero components equal to zero by applying the (integer-valued) 
$T$-duality transformations, as we explain in the following.  We consider 
only the relevant 2-dimensional part of the D$p$-brane worldvolume associated 
with non-zero components $B_{p-1,p}=-B_{p,p-1}=n_{p-1}/N_p$ of the $B$-field, 
i.e. we consider the $O(2,2,{\bf Z})$ $T$-duality transformation.  
The first step is to make the $B$-field take an integer value by applying the 
$T$-duality transformation with the $O(2,2,{\bf Z})$ matrix of the form  
(\ref{bssch}).  We can achieve this, for example, by choosing the entries of 
the $GL(2,{\bf Z})$ matrix $A$ in Eq. (\ref{bssch}) to be integer multiples 
of $N_p$.  The second step is to transform away the resulting integer valued 
non-zero $B$-field components by applying the integer ``$\Theta$''-parameter 
shift $T$-duality transformation with the transformation matrix of the 
form (\ref{thetatran}).  We choose the $2\times 2$ antisymmetric matrix 
$\Theta$ in Eq. (\ref{thetatran}) to be the negative of the $B$-field 
transformed through the first step.  We note that the $T$-duality 
transformations that we applied in the above steps do not change the 
worldvolume dimensionality of the D-brane.  So, we have related a D$p$-brane 
system with constant rank 2 $B$-field to a D$p$-brane system without 
$B$-field.  This result implies the equivalence between the NCYM and CYM 
associate with such D-brane systems.  This result also explains the 
microscopic origin of the equivalence between the thermodynamics of 
the supergravity solutions for D$p$-branes with nonzero $B$-field and those  
with vanishing $B$-field \cite{mar,aos,br,co1,ho,co2}.  Namely, the 
microscopic degrees of freedom of such brane configurations responsible for 
the thermodynamics are mapped one-to-one under the $T$-duality transformation 
mentioned above.  

We notice from the above that different D$p$-brane systems with the non-zero 
$B$-field component $B_{p-1,p}=\tan\theta=n_{p-2}/N_p$ with the fixed number 
$N_p$ of D$p$-branes but with the different number densities $n_{p-2}$ of 
D$(p-2)$-branes (and therefore different values of $B_{p-1,p}$ or $\theta$) 
can be mapped under the $T$-duality transformations to the same number of 
D$p$-branes without the $B$-field.  Therefore, the stringy microscopic 
degrees of freedom for these different D-brane systems are in one-to-one 
correspondence under the $T$-duality transformations.  This gives the 
microscopic explanation for the $\theta$ independence of the thermodynamic 
quantities for the nonextreme supergravity solutions for the (D$(p-2)$,D$p$) 
bound states,  which was previously observed in Refs. 
\cite{mar,aos,br,co1,ho}.  In particular, this implies the equivalence of 
the two extreme limits corresponding to the cases $\theta=0$ and $\theta=
\pi/2$ to the case with a finite nonzero $\theta$.  The $\theta=0$ case is 
just the D$p$-branes without $B$-field, i.e. $B_{p-1,p}=\tan\theta=0$.  
In the $\theta=\pi/2$ case (i.e. $B_{p-1,p}=\tan\theta=\infty$), there are 
infinitely many D$(p-2)$-branes per $(2\pi\sqrt{\alpha^{\prime}})^2$ area of 
the $(x_{p-2},x_p)$-plane, i.e. $n_{p-2}=\infty$.  In this case, the second 
term in the second line of the open string boundary condition in Eq. 
(\ref{bdrcndpp2}) dominates, thereby the boundary condition (\ref{bdrcndpp2}) 
becoming that of open strings attached to D$(p-2)$-branes.  From this, one 
can see the equivalence of the system of D$p$-branes with nonzero constant 
rank 2 $B$-field and the system of infinitely many D$(p-2)$-branes densely 
stacked on the 2-dimensional plane in the transverse space (thereby the 
D$(p-2)$-branes becoming delocalized on the 2-plane), which was previously 
conjectured \cite{town,bfss,ikkt,ish}.  

We comment on the decoupling limit of the D-brane worldvolume theories.  
The NCYM decoupling limit is defined as the limit 
in which $\alpha^{\prime}\sim\varepsilon^{1\over 2}\to 0$ and $g_{ij}\sim
\varepsilon\to 0$ such that $G^{ij}$ and $\theta^{ij}$ in Eq. 
(\ref{propdefs}), including $B_{\rm SW}=B/(2\pi\alpha^{\prime})$, are held 
fixed \cite{sw}.  (Note, the difference in the convention of the $B$-field 
$B_{ij}$ in this paper from the one $B_{{\rm SW}\,ij}$ in Ref. \cite{sw}.)  
So, in the NCYM decoupling limit, the $B$-field behaves as 
\begin{equation}
B_{p-1,p}=\tan\theta={{n_{p-2}}\over{N_p}}={\tilde{b}\over\alpha^{\prime}},
\label{dcplbfld}
\end{equation}
with $\alpha^{\prime}\to 0$ and the noncommutative parameter $\tilde{b}$ 
held fixed.  So, in order for the noncommutative effect on the D-brane 
worldvolume to survive in the decoupling limit, $B_{p-1,p}$ and therefore 
the number density $n_{p-2}$ of D$(p-2)$-branes have to go to infinity.  
The total number of the D$(p-2)$-branes is given by $N_{p-2}=n_{p-2}
V_2/(2\pi\sqrt{\alpha^{\prime}})^2$.   Note, the NCYM decoupling limit 
condition on the coordinates $x_{p-1,p}$ of the supergravity solution 
(\ref{sgsol}) is $x_{p-1,p}={\alpha^{\prime}\over\tilde{b}}\tilde{x}_{p-1,p}$ 
such that $\tilde{x}_{p-1,p}$ are held fixed \cite{hi,rm,aos}.  So, the 
total number of D$(p-2)$-branes is reexpressed as 
\begin{equation}
N_{p-2}=\alpha^{\prime}n_{n-2}\tilde{V}_2/(4\pi^2\tilde{b}^2), 
\label{numofdp2}
\end{equation}
where we used the relation $\tilde{V}_2=\left({\alpha^{\prime}\over\tilde{b}}
\right)^2V_2$.  From Eq. (\ref{dcplbfld}), we see that the D$(p-2)$-brane 
number density goes to infinity as $n_{n-2}\sim 1/\alpha^{\prime}$.  So, even 
if the D$(p-2)$-brane number density $n_{p-2}$ diverges in the NCYM decoupling 
limit, the total number $N_{p-2}$ of the D$(p-2)$-branes can be finite, if 
$\tilde{V}_2<\infty$.  As pointed out in the above, in the $n_{p-2}\to\infty$ 
limit the open string boundary condition (\ref{bdrcndpp2}) reduces to that 
for the open string ending on D$(p-2)$-brane, meaning that the long distance 
behavior of the massless bosonic fields in the closed string states 
interacting with such D-brane system, obtained from the boundary state 
through the projection, reproduces that of the supergravity solution for the 
D$(p-2)$-branes.  Therefore, in the NCYM decoupling limit, the system of 
D$p$-branes with rank 2 constant $B$-field reduces to the system of $N_{p-2}$ 
numbers of D$(p-2)$-branes, which are densely packed on the 2-plane associated 
with non-zero components of the $B$-field, implying the equivalence of NCYM in 
$p+1$ dimensions and CYM with the gauge group $U(N_{p-2})$ in $p-1$ dimensions 
\cite{co2}.  The rank $N_{p-2}$ of the gauge group is determined by $N_p$, 
$\tilde{b}$ and $\tilde{V}_2$ through Eqs. (\ref{dcplbfld}) and 
(\ref{numofdp2}).  When $\tilde{V}_{2}=\infty$, the total number $N_{p-2}$ of 
the D$(p-2)$-branes is infinite and therefore NCYM in $p+1$ dimensions is 
equivalent to CYM with the gauge group $U(\infty)$ in $p-1$ dimensions 
\cite{lur}, as pointed out in Ref. \cite{co2}.  The equivalence of the 
various D-brane configurations connected through the $T$-duality 
transformations, on which we elaborated in the previous paragraphs, therefore 
gives a microscopic explanation for the equivalence among the following gauge 
theories: $(i)$ CYM in $p+1$ dimensions associated with the decoupling limit 
of D$p$-branes without $B$-field, $(ii)$ NCYM in $p+1$ dimensions associated 
with the NCYM decoupling limit of D$p$-branes with nonzero constant rank 2 
$B$-field, $(iii)$ CYM with the gauge group $U(N_{p-2})$ or $U(\infty)$ in 
$p-1$ dimensions.

\end{document}